\documentclass[conference]{IEEEtran}
\usepackage{cite}
\usepackage{amsmath,amssymb,amsfonts}
\usepackage{algorithmic}
\usepackage{graphicx}
\usepackage{textcomp}
\usepackage{xcolor}
\usepackage{fancyhdr}
\usepackage[hyphens]{url}

\usepackage{mathptmx} 
\usepackage[final]{microtype}
\usepackage{flushend}
\usepackage[belowskip=-3pt,aboveskip=3pt]{caption}
\usepackage{amsmath}
\usepackage{tikz}
\usepackage{multicol}
\usepackage[flushleft]{threeparttable}
\usepackage{xspace}
\usepackage{multirow}
\usepackage{rotating}
\usepackage{subfig}
\usepackage{chngpage}
\usepackage{amssymb}
\usepackage{pifont}
\usepackage{xcolor}
\usepackage{mathtools}
\usepackage{float}
\usepackage{color}
\usepackage{comment}
\usepackage[ruled,vlined]{algorithm2e}
\usepackage{enumitem}
\usepackage{amsfonts}
\usepackage{listings}
\usepackage{tabularx}
\PassOptionsToPackage{hyphens}{url}
\usepackage{hhline}
\usepackage{nccmath}

\usepackage[normalem]{ulem}
\useunder{\uline}{\ul}{}

\def\BibTeX{{\rm B\kern-.05em{\sc i\kern-.025em b}\kern-.08em
    T\kern-.1667em\lower.7ex\hbox{E}\kern-.125emX}}

\newcommand{\iscasubmissionnumber}{807}

\newcommand{\bheading}[1]{{\vspace{2pt}\noindent{\textbf{#1}}\hspace{2pt}}}

\newcommand{\cut}[1]{}

\def\authnotes{1}
\newcommand{\authnote}[2]{\ifnum\authnotes=1\begin{quote}\textbf{#1 says:} #2\end{quote}\fi}
\newcommand{\fixme}[1]{\ifnum\authnotes=1\textbf{\textcolor{red}{[FIXME: #1]}}\fi}

\newcommand{\gyrefeq}[1]{Eq (\ref{#1})}
\newcommand{\gyreffig}[1]{Figure~\ref{#1}}
\newcommand{\gyrefsec}[1]{Section~\ref{#1}}
\newcommand{\gyreftbl}[1]{Table~\ref{#1}}

\newcommand{\gyrefapd}[1]{Appendix~\ref{#1}}
\newcommand{\norm}[1]{\left\lVert#1\right\rVert}
\newcommand{\modulenamenobf}{SID}
\newcommand{\modulename}{\modulenamenobf}
\newcommand{\modulefullname}{Smart\-phone Impo\-stor Dete\-ctor}

\newenvironment{packeditemize}{
\begin{list}{$\bullet$}{
\setlength{\labelwidth}{8pt}
\setlength{\itemsep}{0pt}
\setlength{\leftmargin}{\labelwidth}
\addtolength{\leftmargin}{\labelsep}
\setlength{\parindent}{0pt}
\setlength{\listparindent}{\parindent}
\setlength{\parsep}{0pt}
\setlength{\topsep}{3pt}}}{\end{list}}

\makeatletter
\def\@copyrightspace{\relax}
\makeatother

\definecolor{dkgreen}{rgb}{0,0.6,0}
\definecolor{gray}{rgb}{0.5,0.5,0.5}
\definecolor{mauve}{rgb}{0.58,0,0.82}

\fancypagestyle{firstpage}{
  \fancyhf{}

  \fancyhead[C]{\normalsize{ISCA 2020 Submission
      \textbf{\#\iscasubmissionnumber} \\ Confidential Draft: DO NOT DISTRIBUTE}} 
  \fancyfoot[C]{\thepage}
}

\title{Smartphone Impostor Detection with Built-in Sensors and Deep Learning}
\author{
 Guangyuan Hu\\
 Princeton University\\
 Princeton, NJ\\
  \textit{gh9@princeton.edu}
  \and
  Zecheng He\\
   Princeton University\\
 Princeton, NJ\\
  \textit{zechengh@princeton.edu}
  \and
  Ruby Lee\\
   Princeton University\\
 Princeton, NJ\\
  \textit{rblee@princeton.edu}
}

\begin{document}
\maketitle
\pagestyle{plain}


\begin{abstract}

In this paper, we show that sensor-based impostor detection with deep learning can achieve excellent impostor detection accuracy at lower hardware cost compared to past work on sensor-based user authentication (the inverse problem) which used more conventional machine learning algorithms. While these methods use other smartphone users' sensor data to build the (user, non-user) classification models, we go further to show that using only the legitimate user's sensor data can still achieve very good accuracy while preserving the privacy of the user's sensor data (behavioral biometrics). For this use case, a key contribution is showing that the detection accuracy of a Recurrent Neural Network (RNN) deep learning model can be significantly improved by comparing prediction error distributions. This requires generating and comparing empirical probability distributions, which we show in an efficient hardware design. Another novel contribution is in the design of SID (Smartphone impostor Detection), a minimalist hardware accelerator that can be integrated into future smartphones for efficient impostor detection for different scenarios. Our SID module can implement many common Machine Learning and Deep Learning algorithms. SID is also scalable in parallelism and performance and easy to program. We show an FPGA prototype of SID, which can provide more than enough performance for real-time impostor detection, with very low hardware complexity and power consumption (one to two orders of magnitude less than related performance-oriented FPGA accelerators). We also show that the FPGA implementation of SID consumes 64.41X less energy than an implementation using the CPU with a GPU.

\end{abstract}

\section{Introduction}

We rely heavily on our smartphones not only for communication but also to store confidential and private data. Smartphones make it more convenient for us to access our data anywhere, anytime, on mobile devices.

However, smartphone theft is one of the most severe threats to smartphone users, leading to a loss of confidentiality and private data\cite{wang2012smartphonetheft}. Impostors are attackers who take over a smartphone and perform actions allowed for the smartphone owner but not for others. Impostor attacks are critical threats to the confidentiality, privacy and integrity of the secrets and personal information stored in the smartphone and accessible online through the smartphone. For powerful attackers who already know or can bypass the legitimate smartphone user's password or personal identification number (PIN), can a defense-in-depth mechanism be provided to detect impostors quickly before further damage is done? We show how this can be done effectively and implicitly, using an algorithm and architecture approach in this paper.


Implicit impostor detection mechanism built into the smartphone would make stealing smartphones less attractive if the smartphone can detect the impostor, and automatically prevent access to confidential information when an impostor is detected. Unlike explicit authentication with passwords, PINs or fingerprints, we would like to detect a potential impostor by looking for intrinsic differences in smartphone user behavior (also known as behavioral biometrics). For example, is it feasible to detect an impostor as he is just walking with the smartphone?

Our first insight is, the ubiquitous inclusion of motion sensors, e.g. 3-axis accelerometer and the 3-axis gyroscope, in smartphones provide a great opportunity to capture a user's motion patterns. We use data from these two widely-available sensors to characterize the motion of a potential thief currently holding the smartphone, as he is walking away from the theft scene, etc. In the literature, implicit smartphone authentication using sensors is primarily modeled as a binary classification problem using conventional Machine Learning (ML) algorithms like Support Vector Machine (SVM) and Kernel Ridge Regression (KRR)\cite{lee2017implicit}. 

However, these classification approaches require both the legitimate user's sensor data and the sensor data from other users for training. This causes serious privacy issues as users must submit their sensitive behavioral data. In this paper, we explicitly consider the trade-off of security and privacy by investigating one-class anomaly detection for impostor detection. We show that using only the legitimate user's sensor data, we can still achieve very good accuracy while preserving the privacy of the user's sensor data, i.e. the user's behavioral biometrics. 

Our intuition is to build a Recurrent Neural Network (RNN) based deep learning (DL) model that can represent the normal user's behavior. A large deviation of the observed behavior from the model's prediction indicates that the smartphone is used by an impostor. Different from previous work on using RNN for anomaly detection, we show that the detection accuracy of an RNN model can be significantly improved by generating and comparing empirical prediction error distributions (PEDs) rather than just comparing it with a threshold. 

To reduce the attack surface and reduce the cost of impostor detection, we design a small and energy-efficient hardware module for impostor detection. Unlike previous work on ML/DL accelerators whose goal is to get the maximum performance from one or a few specific ML or DL algorithms (e.g. Convolutional Neural Networks ), our goal is not to maximize performance, but to provide \textit{sufficient performance at low cost and low power consumption}. Likewise, our goal is also different from past work that focuses on minimizing power consumption -- rather, we consider a broader goal of trade-offs in security, usability and privacy, with execution time, memory and energy consumption. 

We design \modulefullname~(\modulename) to show that we can efficiently and effectively solve a real security problem with small hardware footprint and energy consumption. 
\modulename~reuses functional units where possible to reduce size and cost. Yet it is also scalable for higher performance if more parallel datapath tracks are implemented. It is designed to support not only the best deep learning algorithms we found for impostor detection in both user scenarios (with and without other users' data for training), but can also support other Machine Learning algorithms, calculations of empirical probability distributions and statistical tests. Programmability support provides flexibility in the choice of algorithms and trade-offs in security, privacy, usability and cost.

Our key contributions are:
\begin{packeditemize}

\item We conduct a comprehensive comparison of ML and DL based classification and one-class outlier detection algorithms for impostor detection. We explicitly consider the trade-offs among security, privacy and usability. Our models achieve 97-98\% accuracy in impostor detection for the multi-user scenario (2-class models), and up to 92\% accuracy in the single-user scenario (1-class models).

\item We show that the detection accuracy of a Recurrent Neural Network (RNN) deep learning model, e.g. LSTM and GRU, can be significantly improved by comparing prediction error distributions.

\item We propose a hardware module that is versatile, scalable and efficient for performing impostor detection without preprocessing or postprocessing on other devices.

\item We show the trade-off between detection accuracy with performance, storage and energy consumption when ML/DL models are mapped to our~\modulename~module.

\item We show a major difference in size between the~\modulename~module and existing accelerators with a similar RNN target.

\item We show \modulename~uses 64.4X less energy than a platform with a GPU.

\end{packeditemize}

\section{Motivation and Threat Model}

Our threat model includes powerful attackers who can bypass the conventional explicit authentication mechanisms, e.g. personal identification number (PIN), password, voice or fingerprint. The explicit authentication bypassing could occur in different ways. For example, the PIN/password may not be strong enough. The attacker can actively figure out the weak pin/password by guessing or social engineering. Another example is the attacker taking the phone after the legitimate user has entered the secret or biometric for explicit authentication. 

Once the explicit authentication is bypassed, the attacker can find and kill any suspicious impostor detection program running as a software process. Hence, we consider a hardware solution that cannot be disabled in this way. 

We assume the smartphone has common, built-in motion sensors, e.g. the accelerometer and the gyroscope. We assume that the sensor readings are always available. 

Our approach can be used with any explicit user authentication mechanism and can provide an extra layer of security in a defense-in-depth approach. Our detection mechanism does not conflict with the conventional explicit authentication but prevents, as much as possible, the exposure of secret information/services to an impostor even when the explicit authentication is broken. 

While this paper assumes a single legitimate user of a smartphone, our detection methodology can be easily extended to allow multiple legitimate users sharing a smartphone, e.g., members in a family or tight-knit team.

\section{Methodology for impostor detection}

Our proposed impostor detection methodology jointly optimizes the security enhancement and the cost of implementation. We consider the security enhancement part in this section, the cost of implementation in~\gyrefsec{sec_hw} and trade-offs in \gyrefsec{sec_evaluation}. Within security enhancement, our methodology explicitly takes three important factors into consideration: attack detection capability (security), usability and user data privacy of the solution.


\subsection{Security and Usability Trade-off}

A good implicit impostor detection solution needs to both be able to detect suspicious impostors and not affect the legitimate user's utilization. At the center of this trade-off is the selection of an appropriate model for impostor detection. One of our key intuitions and takeaways of our methodology is that choosing the right model is more important for achieving security and performance goals than increasing the model size or adding hardware complexity to accelerate a model.

Previous work on implicit smartphone user authentication mostly leverage the (user, non-user) binary classification techniques \cite{frank2012touchalytics, zheng2014you, lee2017implicit}. This scenario requires training models using both data from the real user and potential attackers and we call it the scenario of impostor detection-as-a-service (IDaaS) or Scenario 1 in this paper. The service provider is a centralized party. All end-users (customers) register their phones to the service provider by providing their patterns on motion recorded by the smartphone. The service provider collects data from all customers, creating a model for each customer using the entire data from all customers. For a specific customer, the data from him/herself are labeled as benign (the user's), while all data from other customers are labeled as malicious (non-user's).

We follow the literature and start with selecting certain classification-based machine learning and deep learning models. The goal is to find the algorithms which give the best accuracy for impostor detection (security) and legitimate user recognition (usability). 

We first investigate three representative ML algorithms: logistic regression (LR), support vector machine (SVM) and kernel ridge regression (KRR). Logistic regression intrinsically provides the probability of a class and thus can be naturally leveraged as the warning score. We evaluate SVM because it is a powerful and commonly used linear model, which can establish a non-linear boundary using the kernel method (\gyrefapd{apd_class_alg}). KRR alleviates over-fitting by penalizing large parameters. It is suitable for learning with limited data and achieves the highest detection rate in the literature \cite{lee2017implicit}. Since these machine learning algorithms require heuristic feature selection, we also try the simplest deep learning model, i.e. multi-layer perceptron (MLP). MLP is a non-linear classifier and can automatically extract intricate patterns from raw data, without heuristic and tedious hand-crafting. Details of these algorithms and their equations are given in~\gyrefapd{apd_class_alg}.

\bheading{Metrics.} In order to quantify both factors, i.e. security and usability, of a given binary classification technique for impostor detection, we consider the metrics: true negative rate (TNR) and true positive rate (TPR). Note that a "positive" outcome is the detection of an impostor, while a "negative" outcome implies the legitimate user is detected.

Usability is represented by TNR, which is the percentage of samples when the real user is correctly recognized. 
Security can be measured using TPR, which is the percentage of samples when a wrong user gets detected and rejected. 
\gyrefeq{eq_sw_metric} gives the formula for TNR and TPR, as well as for the metrics commonly used in comparing the ML/DL models: accuracy, recall, precision and F1.

\begin{equation}
\begin{split}
  TNR &
  = \frac{TN}{TN+FP} \\
  TPR &= 
  = \frac{TP}{TP+FN} \\
  Accuracy &= \frac{TN+TP}{TN+FP+TP+FN}\\
  Recall (R) &= TPR \\
  Precision (P) &= \frac{TP}{TP+FP} \\ 
  F1 score &= \frac{2 \times Recall \times Precision}{Recall+Precision}
\end{split}
\label{eq_sw_metric}
\end{equation}



\subsection{Privacy and usability trade-off}

The aforementioned binary classification machine learning approaches can only be applied to the IDaaS scenarios where the data from other users are available. Unfortunately, many smartphone users do not want to share their sensor data for privacy reasons and for fear of being attacked in the network or the cloud. For data privacy concerns \cite{fbleak, googleleak} about users' behavioral biometric data, we need to consider another important scenario where the smartphone user only has his/her own data for training. Therefore, the impostor detection in this scenario can be formalized as a one-class classification, i.e. outlier detection problem. We call this local anomaly detection (LAD) or Scenario 2 in this paper.

Although most of the previous work of continuous smartphone authentication concentrate on binary classifications, there exists some preliminary pioneering work on one-class classification \cite{kumar2018continuous, kazachuk2016one, liatsis2017one}. However, none of these work leverage deep learning techniques on one-class smartphone authentication. 


We consider three representative algorithms to deal with the lack of positive (non-user) training data, i.e. One-Class SVM (OCSVM), Long Short-Term Memory (LSTM) and Gated Recurrent Unit (GRU). We propose enhancing the deep learning models, LSTM and GRU, with the comparison of reference and actual Prediction Error Distributions (PED's). We show in \gyrefsec{sec_alg_eval} that generating and comparing the prediction error distributions is the key to a successful detection for this Scenario 2. The ML/DL prediction algorithm appears to be only of secondary importance, and does not even have to be very accurate.
 
\bheading{OCSVM.} OCSVM is an extension of normal SVM, by separating all the data points from the origin in the feature space and maximizing the distance from this hyperplane to the origin. Intuitively, the OCSVM looks for the minimum support of the normal data, and recognizes points outside this support as anomalies. The kernel method (\gyrefapd{apd_class_alg}) can also be applied to establish a non-linear boundary.

\bheading{LSTM.} Different from the above discussed stateless models (SVM, KRR, OCSVM, etc.), the LSTM model has two hidden states ($h_t$ and $c_t$) which can remember the previous input information.
When being used to model temporal sequences, an LSTM cell updates its hidden states ($h_{t}$, $c_{t}$) for each input time-frame using the previous states ($h_{t-1}$, $c_{t-1}$) and the current input $x_t$ as described in~\gyrefeq{eq_lstm}. Three control gates, the forget gate $f_t$, the input gate $i_t$ and the output gate $o_t$, are used to determine how much of the old states are preserved. This significantly prevents the gradient vanishing problem in training deep neural networks. In~\gyrefeq{eq_lstm}, the W's and U's are weight matrices, and the b's are bias vectors. 

{\baselineskip=0pt
\begin{equation}
\label{eq_lstm}
\begin{split}
  &cand_t = tanh(W_c \times x_{t} + U_c \times h_{t-1} + b_c) \\
  &f_t = \sigma (W_f \times x_{t} + U_f \times h_{t-1} + b_f) \\
  &i_t = \sigma (W_i \times x_{t} + U_i \times h_{t-1} + b_i) \\
  &o_t = \sigma (W_o \times x_{t} + U_o \times h_{t-1} + b_o) \\
  &c_t = f_t \odot c_{t-1} + i_t \odot cand_t \\
  &h_t = o_t \odot tanh(c_t)
\end{split}
\end{equation}
}
\bheading{GRU.} We replace the cell model with a more light-weight Gated Recurrent Unit (GRU) model, described in \gyrefeq{eq_gru}.
{\baselineskip=0pt
\begin{equation}
\label{eq_gru}
\begin{split}
  &z_t = \sigma (W_z \times x_{t} + U_z \times h_{t-1} + b_z) \\
  &r_t = \sigma (W_r \times x_{t} + U_r \times h_{t-1} + b_r) \\
  &h_t = (1-z_t) \odot h_{t-1} + z_t \odot \sigma (W_h x_t + U_r (r_t \odot h_{t-1}) + b_h) \\
\end{split}
\end{equation}
}

We use an LSTM or GRU model as an outlier detector \cite{malhotra2015long}, by training it to predict the next sensor reading, and investigate the prediction errors. The intuition is, an LSTM model trained on only the normal user's data predicts better for his/her behavior than the other users' (potentially an impostor) behavior. The deviation of the real monitored behavior from the predicted behavior indicates the anomalies. Typically, a threshold value is used to decide if the prediction error is normal or not. A key contribution we make is to show that the choice of deep learning models, e.g. LSTM or GRU, is not the most important factor for good performance, if Prediction Error Distributions are leveraged. 

\bheading{LSTM/GRU + Prediction Error Distribution (PED).} Our intuition is that a single prediction error may significantly vary, but the probability distribution of the errors is stable. Therefore, comparing the difference between the observed PED and a reference PED from the real user's validation data is more stable than comparing the average prediction error.

As we do not need to assume the prior distribution of PED, non-parametric tests are powerful tools to determine if two distributions are the same. The Kolmogorov-Smirnov (KS) test is a statistical test that determines whether two i.i.d sets of samples are from the same distribution. The KS statistic for two sets with $n$ and $m$ samples is:
\begin{align}\label{equ:KSstatistic}
D_{n,m} = sup_{x} |F_{n}(x)-F_{m}(x)|
\end{align}

where $F_{n}$ and $F_{m}$ are the empirical distribution functions of two sets of samples respectively, i.e. $F_{n}(t)=\frac{1}{n}\sum_{i=1}^{n}1_{x_i\leq t}$, and $sup$ is the supremum function. The null hypothesis that the two sets of samples are i.i.d. sampled from the same distribution, is rejected at level $\alpha$ if:
{\small
\begin{align}\label{equ:KStestinequ}
D_{n,m}>c(\alpha)\sqrt{\frac{n+m}{nm}}
\end{align}
}%
where $c(\alpha)$ is a pre-calculated value and can be found in the standard KS test lookup table.

\subsection{Experimental Settings}

We evaluate the models for impostor detection using the UCI \cite{Dua:2017} Human Activities and Postural Transitions (HAPT) dataset \cite{reyes2016transition}. The HAPT dataset contains smartphone sensor readings. The smartphone is worn on the waist of a group of 30 participants of various ages from 19 to 48. The participants were asked to perform six activities including standing, sitting, walking, etc. and the transitions between them. The sensor readings are collected and labeled with both the activity and user ID. Each reading consists of the 3-axial measurements of both the linear acceleration and angular velocity, so it could be treated as a 6-element vector. The sensors are sampled at 50Hz. We use the WALK dataset in HAPT to determine the feasibility of user versus impostor classification using the two most common built-in motion sensors and just one common type of motion.

We select 25 out of the 30 users in the HAPT dataset as the registered users while the other 5 users act as unregistered users. For each registered user, models are trained using his/her data and randomly picked sensor data of the other 24 registered users. In the training process, all the data from the other 24 registered users are labeled as positive for impostor detection and not the correct user. The unregistered users are used in testing to examine whether unseen attackers can be successfully detected.

As \cite{lee2017implicit} reports very high accuracy for recognizing users using sensor data, we implement the same KRR model which first computes 14 features: for each of the 2 sensors (accelerometer and gyroscope), 4 are common statistics like min, max, mean and standard deviation, in the time domain and 3 are frequency domain features.

\subsection{Security vs Usability Evaluation}
\label{sec_alg_eval}

We show the results of two-class classification in Table \ref{tbl_accuracy_binary}. We choose a window size of 64 \footnote{It is easier to extract FFT features if the window size is a power of 2.} sensor readings because this window size covers 1-2 full human steps given the sensors are sampled at 50 Hz. We observe that the SVM model performs the best for all the metrics. Surprisingly, it also performs better than KRR with manually selected features.
However, the simple deep learning model, MLP, performs almost as well, and costs less; SVM requires more storage and execution time than MLP for a marginal 1-2\% of accuracy increase (see left half of \gyreffig{img_hw_perf} and \gyreffig{img_hw_mem} in Section \ref{sec_evaluation}). These two figures will also show that, MLP with 2 hidden levels (MLP-200-100) is better, from the memory and execution-time performance perspective, than MLP with one hidden layer but more nodes (MLP-500), while both have 97\% accuracy.


\begin{table}[ht]
\centering
\resizebox{\columnwidth}{!}{
\begin{tabular}{|l|c|c|c|c|c|}
\hline
\multirow{2}{*}{Models}       & \multicolumn{5}{c|}{All Users}                                                                                                                                                                                                         \\ \cline{2-6} 
                              & \begin{tabular}[c]{@{}c@{}}TNR\\ (\%)\end{tabular} & \begin{tabular}[c]{@{}c@{}}TPR/Recall\\ (\%)\end{tabular} & \begin{tabular}[c]{@{}c@{}}Accuracy\\ (\%)\end{tabular} & P                            & F1                           \\ \hline
LR                            & 80.77                                              & 66.05                                                     & 73.41                                                   & 0.77                         & 0.69                         \\ \hline
KRR                           & 88.91                                              & 82.66                                                     & 85.78                                                   & 0.87                         & 0.83                         \\ \hline
\textit{\textbf{SVM}}         & {\ul \textit{\textbf{99.26}}}                      & {\ul \textit{\textbf{97.57}}}                             & {\ul \textit{\textbf{98.42}}}                           & {\ul \textit{\textbf{0.99}}} & {\ul \textit{\textbf{0.98}}} \\ \hline
MLP-50                        & 98.31                                              & 92.70                                                     & 95.51                                                   & 0.98                         & 0.94                         \\ \hline
MLP-100                       & 98.60                                              & 94.65                                                     & 96.63                                                   & 0.98                         & 0.96                         \\ \hline
MLP-200                       & 98.41                                              & 95.72                                                     & 97.06                                                   & 0.98                         & 0.97                         \\ \hline
\textit{\textbf{MLP-500}}     & {\ul \textit{\textbf{98.68}}}                      & {\ul \textit{\textbf{95.47}}}                             & {\ul \textit{\textbf{97.07}}}                           & {\ul \textit{\textbf{0.99}}} & {\ul \textit{\textbf{0.96}}} \\ \hline
MLP-50-25                     & 98.13                                              & 94.49                                                     & 96.31                                                   & 0.98                         & 0.96                         \\ \hline
MLP-100-50                    & 98.44                                              & 95.45                                                     & 96.95                                                   & 0.98                         & 0.96                         \\ \hline
\textit{\textbf{MLP-200-100}} & {\ul \textit{\textbf{98.47}}}                      & {\ul \textit{\textbf{95.72}}}                             & {\ul \textit{\textbf{97.10}}}                           & {\ul \textit{\textbf{0.98}}} & {\ul \textit{\textbf{0.97}}} \\ \hline
\end{tabular}
}
\caption{Impostor detection in the IDaaS secnario, using binary classification ML and DL models, achieves 97\%-98\% accuracy.}
\label{tbl_accuracy_binary}
\end{table}

We use the same experimental settings on HAPT dataset for the evaluation of one-class outlier detection approaches. We train an OCSVM model with a fixed window size of 64. The latency for detection is a multiple of 20-ms (e.g.,64x, or 1.28 seconds) which is comparable to the time for an impostor to act. The LSTM or GRU model makes predictions for every sensor reading in a fixed window of size 200. The classification results of \textbf{LSTM-\#} and \textbf{GRU-\#} (rows 2 through 6 in \gyreftbl{tbl_accuracy_oneclass}, \# is the size of hidden states) are given by comparing the average prediction error with a threshold obtained from the validation set. The PED-enhanced LSTM or GRU leverages the distribution of prediction errors within the same window as LSTM or GRU. We first randomly choose 20 samples of prediction errors from the validation set and use them to represent the reference PEDs. In the testing phase, the prediction error distribution of each testing sample is compared to all the reference distributions. If half of the KS-test-statistics are larger than a commonly used p-value threshold, e.g. 0.05 in our experiment, the current sample is considered as abnormal. 

We show the results of three types of one-class models, i.e. OCSVM, LSTM/GRU, and LSTM or GRU + PED, in Table \ref{tbl_accuracy_oneclass}. The one-class SVM achieves an average accuracy of 69.2\%, thirty percent worse than classification models trained with positive data involved. The standard LSTM or GRU models have an accuracy of $\sim$ 70\%, only slightly better than the one-class SVM model, regardless of the network architecture. On the contrary, there exists a significant improvement if PED and statistical KS test are leveraged, e.g. the accuracy of LSTM or GRU models + KS test reach 89.0\% or 92.3\%, respectively. We conclude that, neither the choice of deep learning models, e.g. LSTM or GRU (3.5\% gain), nor the design of network architectures (1.5\% gain) is the first essential factor for good performance. However, it is the Prediction Error Distributions and KS test that provide the significant increase in detection capability, i.e. +19.2\% for the best LSTM and +19.4\% for the best GRU. Without loss of generality, we use the more conservative numbers for LSTM in the rest of the paper (GRU will be slightly better).

Once PED and KS are used, we observe that the PED-LSTM-200-OCSVM and PED-LSTM-200-VOTE tradeoff between security and usability, respectively. The PED-LSTM-200-OCSVM achieves a high TPR of 92.25\% but a lower TNR of 74.82\%, i.e. a ``strict'' detection providing good security against impostors but can erroneously reject the real user. The voting-based method, i.e. PED-LSTM-200-VOTE, achieves a high TNR of 94.62\% but a lower TPR of 83.31\%, which means it is a ``lenient'' detection mechanism providing good usability for the correct user but limited security against impostors. Both approaches significantly outperform the non-PED (non-KS) approaches.

Given that adding the KS test is critical to improving the detection accuracy in this scenario, we provide the hardware support for generating empirical PEDs and computing the KS statistic in~\gyrefsec{sec_op_kstest}. 

\begin{table}[ht]
\centering
\resizebox{0.48\textwidth}{!}{
\begin{tabular}{|l|c|c|c|c|c|}
\hline
\multirow{2}{*}{Models}                                       & \multicolumn{5}{c|}{All Users}                                                                                                                                                         \\ \cline{2-6} 
                                                              & \begin{tabular}[c]{@{}c@{}}TNR\\ (\%)\end{tabular} & \begin{tabular}[c]{@{}c@{}}TPR/Recall\\ (\%)\end{tabular} & \begin{tabular}[c]{@{}c@{}}Accuracy\\ (\%)\end{tabular} & P    & F1   \\ \hline
OCSVM                                                         & 64.24                                              & 74.19                                                     & 69.22                                                   & 0.59 & 0.65 \\ \hline
LSTM-50                                                       & 72.43                                              & 67.04                                                     & 69.74                                                   & 0.57 & 0.60 \\ \hline
LSTM-100                                                      & 72.20                                              & 69.27                                                     & 70.73                                                   & 0.58 & 0.62 \\ \hline
LSTM-200                                                      & 67.88                                              & 71.60                                                     & 69.74                                                   & 0.58 & 0.62 \\ \hline
LSTM-500                                                      & 67.57                                              & 74.42                                                     & 70.99                                                   & 0.60 & 0.65 \\ \hline
GRU-200                                                       & 69.12                                              & 76.68                                                     & 72.9                                                    & 0.63 & 0.68 \\ \hline
\begin{tabular}[c]{@{}l@{}}PED-LSTM-\\ 200-OCSVM\end{tabular} & 74.82                                              & {\ul \textit{\textbf{92.25}}}                             & 83.53                                                   & 0.80 & 0.84 \\ \hline
\begin{tabular}[c]{@{}l@{}}PED-GRU-\\ 200-OCSVM\end{tabular}  & 78.47                                              & {\ul \textit{\textbf{93.58}}}                             & 86.03                                                   & 0.82 & 0.86 \\ \hline
\begin{tabular}[c]{@{}l@{}}PED-LSTM-\\ 200-Vote\end{tabular}  & {\ul \textit{\textbf{94.62}}}                      & 83.31                                                     & 88.97                                                   & 0.91 & 0.83 \\ \hline
\begin{tabular}[c]{@{}l@{}}PED-GRU-\\ 200-Vote\end{tabular}   & {\ul \textit{\textbf{96.86}}}                      & 87.79                                                     & {\ul \textit{\textbf{92.33}}}                           & 0.94 & 0.88 \\ \hline
\end{tabular}
}
\caption{Impostor detection accuracy in the LAD secnario, using one-class models}
\label{tbl_accuracy_oneclass}
\end{table}

Note that our impostor detection methodology is an extra layer in a defense-in-depth approach that does not conflict with explicit authentication using PIN, voice or fingerprint. Very high sensitivity levels (95\%-99\%) are achieved for accuracy, security (TPR) and usability (TNR) when SVM or MLP models are used in the IDaaS scenario. If sensor data privacy is required, the accuracy is lower but still acceptable: security (92\%-93\%) or usability (94\%-96\%) can be achieved using our enhanced LSTM or GRU models, depending on one's needs. Although the accuracy is not perfect, it is actually comparable to the state-of-the-art one class smartphone authentication using various handcrafted features and complex model fusion \cite{kumar2018continuous} in the literature.


\section{\modulenamenobf~Hardware Module}
\label{sec_hw}

Our goal is to design a small but versatile and program\-mable hardware module that can be integrated into a smartphone to perform impostor detection, without needing compu\-tation on another processor or accelerator. Design goals for our \modulefullname~(\modulename) are:
\begin{packeditemize}
\item {Flexibility for different ML/DL models and trade-offs of security, usability, privacy, execution time, storage and energy consumption}, 
\item {Scalability for more performance}, 
\item Reduced memory storage and access,
\item Minimized energy consumption, and 
\item {Reduced attack surface for better security}.

\end{packeditemize}

While our primary goal is to design \modulename~to be able to perform the best algorithms for impostor detection, namely, MLP and SVM for the IDaaS binary classification scenario, and one-class PED-LSTM/GRU for the local detection scenario, we also want it to be flexible enough to support other ML/DL algorithms as well, possibly for future security needs. For performance scalability, we design \modulename~to allow more parallel tracks to be implemented, if desired. An innovative aspect of our design is that the \modulename~macro instructions implementing the selected ML/DL algorithm do not even have to be changed when the number of parallel tracks is increased. 
We also design \modulename~to reduce memory traffic by reusing operands and having local storage in the execution stage.
To be feasible for implementation in a smartphone, we design \modulename~for reduced energy consumption to alleviate smartphone battery usage. To reduce the attack surface, \modulename~should be able to support detection without subsequent processing on another device like the CPU. For example, our \modulename~also does statistical testing, which is important to enhance ML/DL models. This includes collecting and comparing empirical probability distributions.



\subsection{Architecture Overview}
\label{sec_arch_overview}


For flexibility, we want \modulename~to be able to support different ML/DL algorithms, for different security/usability/privacy/cost trade-offs, as well as future security needs.



\begin{figure}[ht]
    \centering
    \includegraphics[width=\linewidth]{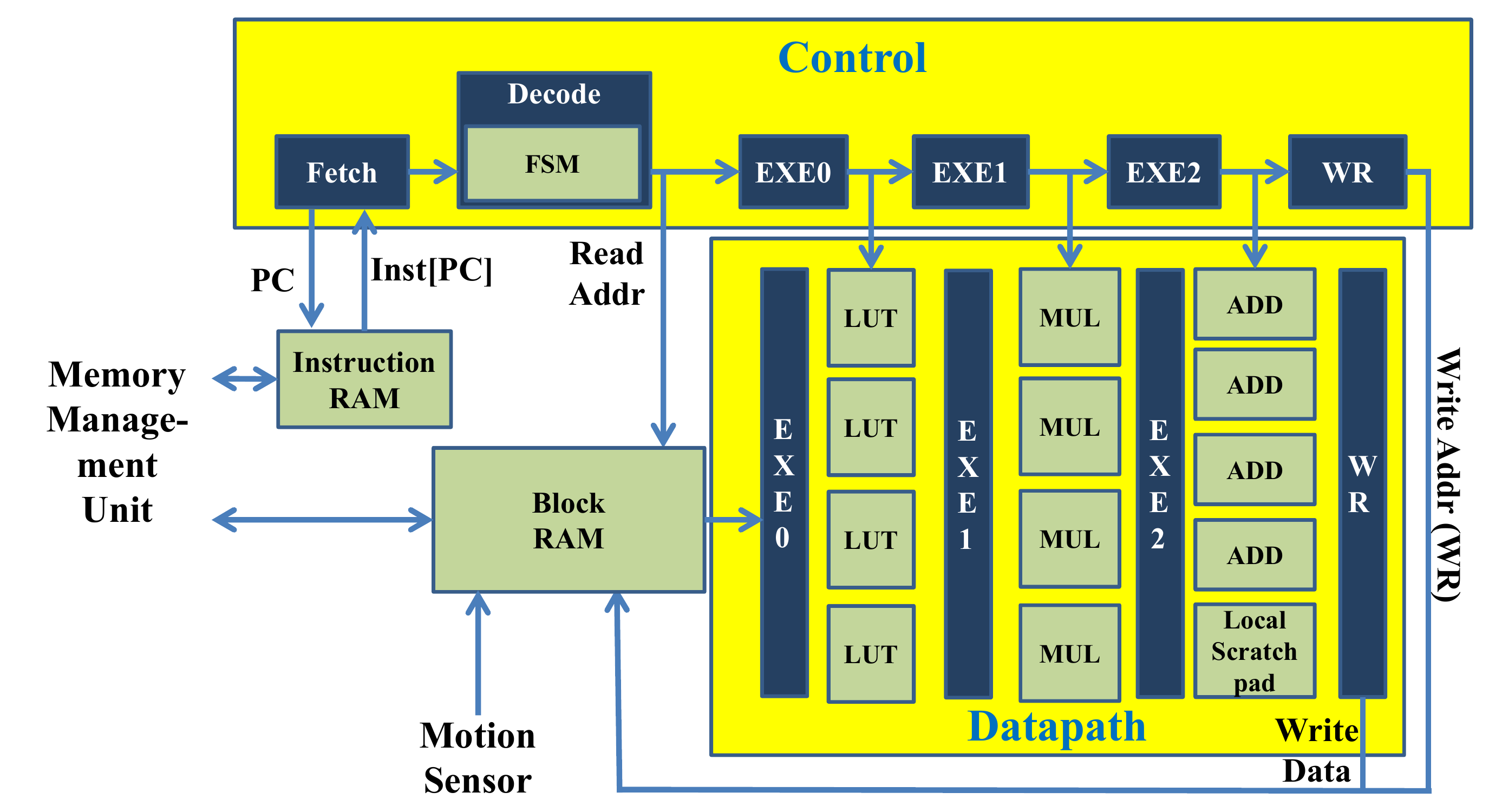}
    \caption{A \modulename~module implementing vector and matrix operations defined in \gyreftbl{tbl_primitives}}
    \label{img_blockdiagram}
\end{figure}

An overview of \modulename~is shown in \gyreffig{img_blockdiagram}. Unlike vector machines with vector registers of fixed length, we use memory (BRAM in the FPGA) to store the vector or matrix operands and results. The memory controlled by macro instructions in \modulename~is more efficient since it does not have fixed vector lengths and operates seamlessly with our automatic determination of the number of times to execute a vector or matrix operation. We also have a small local scratchpad memory for faster access to intermediate results during a macro operation. The scratchpad is used to reduce the memory traffic for matrix-vector multiplication and other vector operations (\gyrefsec{sec_parallel_operation}).

We propose a minimal implementation of a \modulename~module with four parallel tracks in the datapath, as shown in \gyreffig{img_blockdiagram}. Each track consists of a Look-up Table (LUT), a Multiplier and an Adder, which are put into three consecutive EXEcution stages. \gyrefsec{sec_macro_iteration} shows how \modulename~is programmed, how it controls the execution of macro instructions and how \modulename~is scalable for loops. We describe the operations implemented with maximal functional unit reuse and the memory access reduction in~\gyrefsec{sec_parallel_operation}. An efficient implementation of empirical probability distribution generation and comparison is given in~\gyrefsec{sec_op_kstest}.

\bheading{Integration in smartphone hardware system.} Compared to conventional processing which requires the sensor data to be saved in main memory and then read out by application processors or other computation devices, integrating a detection module closer to the sensors can reduce the attack surface and save the overhead of memory accesses. Modern smartphones have already implemented the interface to write the collected sensor input to a cache memory for efficient signal processing~\cite{hexagon:hotchip2015:codrescu2015architecture}. The \modulename~module can leverage a similar interface (the input "Motion Sensors" in~\gyreffig{img_blockdiagram}). A valid incoming sensor input can reset the program counter of \modulename~to the beginning of the detection program. 

\begin{table}[ht]
\centering
\resizebox{0.48\textwidth}{!}{
\begin{tabular}{crcrcrcrcrcr}
\multicolumn{1}{l}{\textbf{127}} & \textbf{124} & \multicolumn{1}{l}{\textbf{123}} & \textbf{110} & \multicolumn{1}{l}{\textbf{109}} & \textbf{96} & \multicolumn{1}{l}{\textbf{95}} & \textbf{64} & \multicolumn{1}{l}{\textbf{63}} & \textbf{32} & \multicolumn{1}{l}{\textbf{31}} & \textbf{0} \\ \hline
\multicolumn{2}{|c|}{Mode}                      & \multicolumn{2}{c|}{Length}                     & \multicolumn{2}{c|}{Width}                     & \multicolumn{2}{c|}{Addr\_x}                  & \multicolumn{2}{c|}{Addr\_y}                  & \multicolumn{2}{c|}{Addr\_z}                 \\ \hline
\multicolumn{2}{|c|}{4 bits}                    & \multicolumn{2}{c|}{14 bits}                    & \multicolumn{2}{c|}{14 bits}                   & \multicolumn{2}{c|}{32 bits}                  & \multicolumn{2}{c|}{32 bits}                  & \multicolumn{2}{c|}{32 bits}                 \\ \hline
\end{tabular}
}
\caption{\modulename~Macro Instruction with Automatic Scalable FSM Control
}
\label{tbl_inst_fmt}

\end{table}

\subsection{Scalable FSM Control with Macro Operations}
\label{sec_macro_iteration}

We provide a scalable programming interface to allow ML/DL models with different operations and sizes. The software in \modulename~is a sequence of macro-instructions, which encode a flexible number of iterations for an entire vector or matrix operation.
Each macro instruction initializes the finite state machine (FSM) state of the control unit to indicate the number of iterations of the specified operation. Each cycle, the FSM (in decode stage in~\gyreffig{img_blockdiagram}) updates the number of uncomputed iterations, according to the number of parallel tracks, to decide when a macro-instruction finishes (details in the following paragraphs).
This means the same code can run on the \modulename~hardware modules with a different number of parallel tracks without modification.

The format of a \modulename~macro instruction is shown in~\gyreftbl{tbl_inst_fmt}. The \textbf{Mode} field specifies one of the operation modes in~\gyreftbl{tbl_primitives}, and the~\textbf{Length} and~\textbf{Width} fields initialize the control FSM and indicate when the execution of the macrio instruction finishes. 
The control FSM has three state registers,~\textbf{reg\_le\-ngth},~\textbf{reg\_wid\-th} and~\textbf{reg\_wid\-th\_copy}, to track the number of iterations that have not been finished and automatically decide when to stop the FSM and move to the next instruction (which initiates a new FSM). The FSM can be configured by instructions in two ways: the one-dimension iteration and the matrix-vector iteration.

\begin{table*}[ht]
\centering
\resizebox{\textwidth}{!}{

\begin{tabular}{|l|l|c|c|c|c||c|c|c|c|c|c|c|}
\hline
\multirow{2}{*}{\textbf{\begin{tabular}[c]{@{}l@{}}Operation\\ Modes\end{tabular}}} & \multirow{2}{*}{\textbf{Description}}                                                                           & \multicolumn{4}{c||}{\textbf{IDaaS}}                                                                                                                                             & \multicolumn{6}{c|}{\textbf{LAD}}                                                                                                                                                                                                                                                                                         & \multirow{2}{*}{\textbf{\begin{tabular}[c]{@{}c@{}}Support\\ Status\end{tabular}}} \\ \cline{3-12}
                                                                                    &                                                                                                                 & \textbf{LR}               & \multicolumn{1}{c|}{\textbf{KRR}} & \textbf{\begin{tabular}[c]{@{}c@{}}SVM w/\\ Gaussian\\ Kernel\end{tabular}} & \multicolumn{1}{c||}{\textbf{MLP}} & \textbf{\begin{tabular}[c]{@{}c@{}}OCSVM w/\\ Gaussian\\ Kernel\end{tabular}} & \textbf{LSTM}             & \multicolumn{1}{c|}{\textbf{GRU}} & \textbf{KS-test}          & \textbf{\begin{tabular}[c]{@{}c@{}}PED-\\ LSTM/GRU\\ -OCSVM\end{tabular}} & \textbf{\begin{tabular}[c]{@{}c@{}}PED-\\ LSTM/GRU\\ -Mvote\end{tabular}} &                                                                                    \\ \hline
\textbf{Vadd}                                                                       & \textbf{\begin{tabular}[c]{@{}l@{}}Element-wise addition\\ of two vectors\end{tabular}}                         & \checkmark & \checkmark         & \checkmark                                                   & \checkmark         & \checkmark                                                     & \checkmark & \checkmark         & \checkmark & \checkmark                                             & \checkmark                                             & Yes                                                                                \\ \hline
\textbf{Vsub}                                                                       & \textbf{\begin{tabular}[c]{@{}l@{}}Element-wise subtraction\\ of two vectors\end{tabular}}                      &                           &                                   & \checkmark                                                   &                                   & \checkmark                                                     & \checkmark & \checkmark         & \checkmark & \checkmark                                             & \checkmark                                             & Yes                                                                                \\ \hline
\textbf{Vmul}                                                                       & \textbf{\begin{tabular}[c]{@{}l@{}}Element-wise multiplication\\ of two vectos\end{tabular}}                    &                           & \checkmark         & \checkmark                                                   &                                   & \checkmark                                                     & \checkmark & \checkmark         & \checkmark & \checkmark                                             & \checkmark                                             & Yes                                                                                \\ \hline
\textbf{Vsgt}                                                                       & \textbf{\begin{tabular}[c]{@{}l@{}}Element-wise set-greater-than\\ of two vectors\end{tabular}}                 & \checkmark & \checkmark         & \checkmark                                                   & \checkmark         & \checkmark                                                     & \checkmark & \checkmark         &                           & \checkmark                                             & \checkmark                                             & Yes                                                                                \\ \hline
\textbf{Vsig}                                                                       & \textbf{Sigmoid function of a vector}                                                                           &                           & \checkmark         &                                                                             & \checkmark         &                                                                               & \checkmark & \checkmark         &                           & \checkmark                                             & \checkmark                                             & Yes                                                                                \\ \hline
\textbf{Vtanh}                                                                      & \textbf{Tanh function of a vector}                                                                              &                           &                                   &                                                                             & \checkmark         &                                                                               & \checkmark & \checkmark         &                           & \checkmark                                             & \checkmark                                             & Yes                                                                                \\ \hline
\textbf{Vexp}                                                                       & \textbf{Exponential function of a vector}                                                                       &                           &                                   & \checkmark                                                   &                                   & \checkmark                                                     &                           &                                   &                           & \checkmark                                             &                                                                       & Yes                                                                                \\ \hline
\textbf{Mvmul}                                                                      & \textbf{\begin{tabular}[c]{@{}l@{}}Multiplication of a\\ matrix and a vector\end{tabular}}                      & \checkmark & \checkmark         & \checkmark                                                   & \checkmark         & \checkmark                                                     & \checkmark & \checkmark         &                           & \checkmark                                             & \checkmark                                             & Yes                                                                                \\ \hline
\textbf{VSsgt}                                                                      & \textbf{\begin{tabular}[c]{@{}l@{}}Set-greater-than to compare a\\ scalar and a vector's elements\end{tabular}} &                           &                                   &                                                                             &                                   &                                                                               &                           &                                   & \checkmark & \checkmark                                             & \checkmark                                             & Yes                                                                                \\ \hline
\textbf{Vmaxabs}                                                                    & \textbf{\begin{tabular}[c]{@{}l@{}}Find the maximum absolute\\ value of a vector\end{tabular}}                  &                           & \checkmark         &                                                                             &                                   &                                                                               &                           &                                   & \checkmark & \checkmark                                             & \checkmark                                             & Yes                                                                                \\ \hline
\textbf{Vsqnorm}                                                                    & \textbf{Squared L2-norm of a vector}                                                                            &                           &                                   & \checkmark                                                   &                                   & \checkmark                                                     & \checkmark & \checkmark         &                           & \checkmark                                             & \checkmark                                             & Yes                                                                                \\ \hline
\hhline{|=|=|=|=|=|=|=|=|=|=|=|=|=|}
\textbf{Vargmax}                                                                    & \textbf{\begin{tabular}[c]{@{}l@{}}Find the index of the\\ maximum in a vector\end{tabular}}                    &                           & \checkmark         &                                                                             &                                   &                                                                               &                           &                                   &                           &                                                                       &                                                                       & No                                                                                 \\ \hline
\textbf{Vmin}                                                                       & \textbf{Find the minimum in a vector}                                                                           &                           & \checkmark         &                                                                             &                                   &                                                                               &                           &                                   &                           &                                                                       &                                                                       & No                                                                                 \\ \hline
\textbf{Vmax2}                                                                      & \textbf{\begin{tabular}[c]{@{}l@{}}Find the second largest\\ number in a vector\end{tabular}}                   &                           & \checkmark         &                                                                             &                                   &                                                                               &                           &                                   &                           &                                                                       &                                                                       & No                                                                                 \\ \hline
\textbf{VFFT}                                                                       & \textbf{\begin{tabular}[c]{@{}l@{}}Compute the Fourier\\ transform of a vector\end{tabular}}                    &                           & \checkmark         &                                                                             &                                   &                                                                               &                           &                                   &                           &                                                                       &                                                                       & No                                                                                 \\ \hline
\textbf{Vsqrt}                                                                      & \textbf{\begin{tabular}[c]{@{}l@{}}Compute the square root\\ of each element in a vector\end{tabular}}          &                           & \checkmark         &                                                                             &                                   &                                                                               &                           &                                   &                           &                                                                       &                                                                       & No                                                                                 \\ \hline
\end{tabular}

}

\caption{Computation primitives needed by different ML/DL models and statistical testing.}
\label{tbl_primitives}
\end{table*}

The one-dimension iteration in a vector operation uses only the \textbf{reg\_length}. The value of \textbf{reg\_length} is initialized by the~\textbf{length} field of the instruction. During execution, \textbf{reg\_length} is decreased every cycle by~\textbf{N(track)}, which is the number of parallel tracks, until \textbf{reg\_length} is no larger than~\textbf{N(track)} and the FSM lets the module fetch the next instruction.

\begin{figure}[ht]
    \centering
    \includegraphics[width=\linewidth]{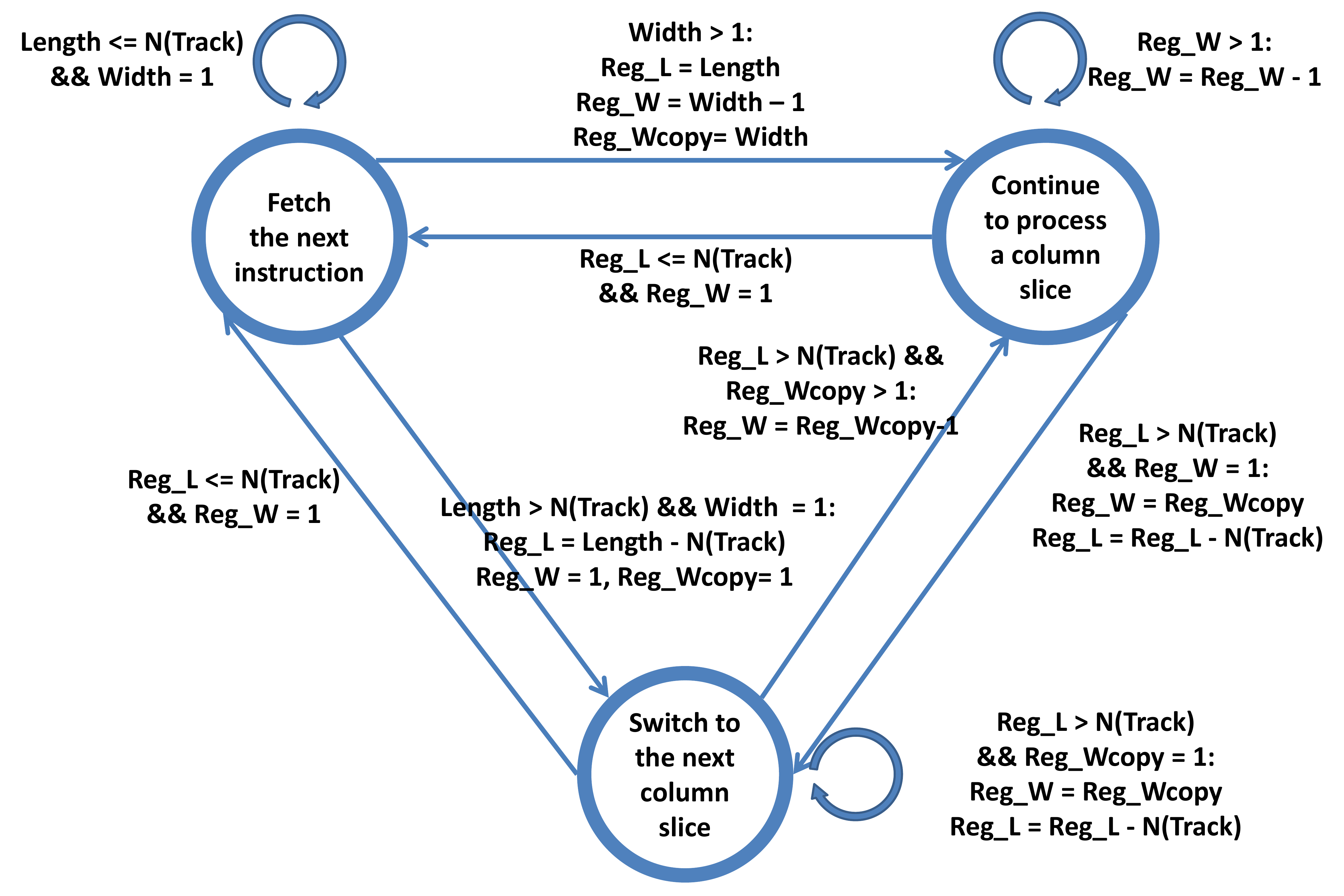}
    \caption{FSM for matrix-vector iteration, e.g. Mvmul. N(track) is the number of parallel tracks.}
    \label{img_fsm_diagram_mvmul}
\end{figure}

The matrix-vector iteration, is used in a tiling manner. It involves all three state registers. 
\gyreffig{img_fsm_diagram_mvmul} shows how the FSM gets updated every cycle in the matrix-vector multiplication mode. When an instruction for matrix-vector operation is fetched, the \textbf{length} field initializes \textbf{reg\_le\-ngth} and the \textbf{wid\-th} field initializes both \textbf{reg\_wid\-th} and \textbf{r\-eg\_wid\-th\_copy}.  
~\textbf{reg\_width}, which gets decremented every cycle, 
controls the downward iteration along a matrix column. When one inner iteration using \textbf{reg\_wid\-th} is finished, \textbf{r\-eg\_wid\-th\_copy} refills \textbf{reg\_width} and \textbf{reg\_le\-ngth} gets decremented by \textbf{N(track)}, which is the number of parallel tracks as defined above. 
This corresponds to moving to the next slice of matrix columns. When the last slice of columns in the matrix is computed, the next instruction can be fetched.

This encoding-control mechanism has the advantages of both scalability and performance. The design is scalable since the hardware is aware of the number of parallel data tracks that are implemented and can perform automatic control of the loop(s). For performance, the control by FSM avoids using branch instructions for frequent jump-backs as needed in general-purpose processors, which can take up a large portion of processor throughput for simple loop bodies.

\subsection{Functional Operations Supported}
\label{sec_parallel_operation}

Our basic hardware module, \modulename, supports all the operations from \textbf{Vadd} through \textbf{Vsqnorm} in~\gyreftbl{tbl_primitives}. 
The instructions from \textbf{Vargmax} to \textbf{Vsqrt} (at the bottom of~\gyreftbl{tbl_primitives}) are needed only for extracting the features, e.g. computing minimum, maximum, square-root, Fourier transform frequency components, for a KRR algorithm~\cite{lee2017implicit}. We decide not to implement these since they introduce significant hardware complexity while achieving lower accuracy than the other algorithms (see \gyreftbl{tbl_accuracy_binary}).

For a detection algorithm, programming \modulename~is directly writing a sequence of instructions corresponding to the operations in the equations for model inference, and specifying the sizes of the matrices and vectors, i.e. model parameters, in the instruction.

We describe briefly the vector and matrix-vector operation modes supported by going down the rows of~\gyreftbl{tbl_primitives}. Key design optimizations are reusing functional units to reduce the hardware cost, automatic FSM operation, tiling and adding the local scratchpad memory to reduce memory accesses.

\bheading{Basic vector-vector operations.}
The addition (\textbf{Vadd}), subtraction (\textbf{Vsub}) and multiplication (\textbf{Vmul}) are straightforward. Vector comparison (\textbf{Vsgt}) is implemented as a set-greater-than (SGT) operation where 
{
\begin{equation}
c[i] = \begin{cases}
    1, & a[i] \geq b[i] \\
    0, & a[i] < b[i]
\end{cases}
, i = 0, ..., len(c)-1
\end{equation}
}
In the~\modulename~module,~\textbf{Vadd},~\textbf{Vsub} and~\textbf{Vsgt} modes are computed in parallel by the 4 parallel tracks shown in~\gyreffig{img_blockdiagram}~in the adder stage (EXE2). The~\textbf{Vmul} mode only uses the multiplier stage (EXE1) for parallel multiplication.

\bheading{Reuse of functional units in vector non-linear functions.}
When computing non-linear functions like sigmoid (\textbf{Vsig}), tanh (\textbf{Vtanh}) and exponential (\textbf{Vexp}), 
we use the look-up table (LUT) implementation to compute the linear approximation. This avoids needing to implement complex non-linear functions, while providing the flexibility of implementing arbitrary non-linear functions by table lookup of the slope and intercept for linear approximation.
An added benefit of our approach over prior work \cite{diannao:Chen:2014:DSH:2541940.2541967} is that we place the LUTs before the multipliers and adders in the three consecutive execution stages so that no extra multipliers or adders are needed.
The ELE0 (LUT) stage of \modulename~outputs a slope, $k(i)$, and an intercept, $b(i)$, for each input value. The interpolation is then computed in the later two stages as $Z[i] = k[i] \times X[i] + b[i]$ with Z[i] being the value of the non-linear function for input X[i].

\bheading{Local scratchpad with tiling and reuse of adders in matrix-vector multiplication.}
\textbf{MVmul} mode computes the product of a weight matrix and an input vector. 
We have two optimizations for \textbf{MVmul} mode.

First, instead of having another adder tree stage to sum the products computed in the EXE1 stage, we put multiplexers before the adders in EXE2 stage to reuse the adders, which saves on hardware cost. 
In the 4-track example, the first three adders sum the four products. The fourth one adds the sum to the partial sum read from the local scratchpad and writes the new partial sum back to the local scratchpad if the computation is not finished. 

\begin{figure}[ht]
    \centering
    \includegraphics[width=\linewidth]{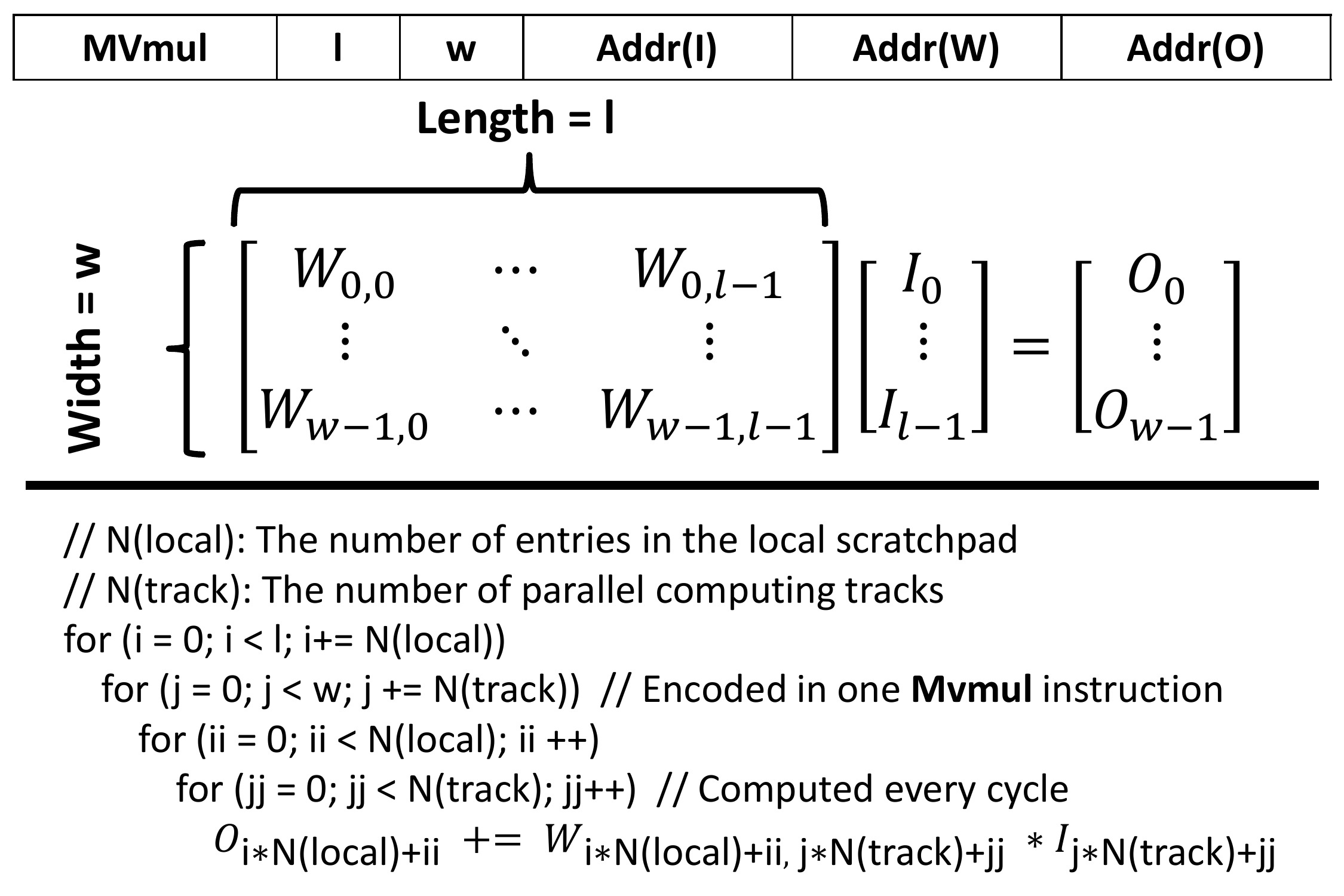}
    \caption{The macro instruction (top) and tiling (bottom) of the matrix-vector multiplication in the \textbf{MVmul} Mode}
    \label{img_matrix}
\end{figure}

Second, we use a local scratchpad memory and loop tiling to save the latency of storing and accessing partial sums from memory and also reduce the memory traffic. A specification of a matrix-vector multiplication is shown in \gyreffig{img_matrix}~followed by the pseudo-code description of tiling. \textbf{N\_track} and \textbf{N\_local} in the pseudo-code are implementation parameters of \modulename~that are provided to the instruction generator. When an output entry of the vector \textbf{O} cannot be computed in one cycle, its partial sum is stored in the local scratchpad memory in the EXE2 stage (see \gyreffig{img_blockdiagram}) so that it does not need to be fetched from memory every cycle. Also, every cycle, \textbf{N\_track} weights (the inner loop of \textbf{jj}) in the same row and their corresponding input elements are multiplied and added to one partial sum. The weights at the same location in the rows below (the loop of \textbf{ii}) will be used in the following cycles, so the input data is reused. The same reuse of the input data is exploited for all columns (the loop of \textbf{j}). A single \textbf{MvMul} instruction can perform all the loops except the first loop of \textbf{i}. When the local scratchpad memory cannot hold all the partial sums, the loop of \textbf{i} is performed using multiple \textbf{MvMul} instructions, each of which corresponds to an FSM in \gyreffig{img_fsm_diagram_mvmul}. 

\bheading{Adding vector-scalar comparison for statistical test.} The vector-scalar comparison mode,~\textbf{VSsgt}, is required by the statistical test support (see~\gyrefsec{sec_op_kstest}). It is computed by setting an output element to one if the corresponding element in the input vector is larger than the other scalar operand, otherwise zero. During execution, the scalar is kept in the pipeline, and the vector elements are streamed in for comparison.

\bheading{Using local scratchpad for fast accesses to intermediate results.}
Besides \textbf{Mvmul}, other operation modes, which need to store intermediate results, may also use the local scratchpad to avoid always reading them from memory outside the \modulename~module. The~\textbf{Vmaxabs} mode compares the absolute values of input elements in EXE2 stage by doing subtraction with the adders. A temporary maximum is stored in the local scratchpad and gets updated every cycle. 
The~\textbf{Vsqnorm} mode computes the squ\-ared L2-\-norm of a vector. 
The scratchpad is also used in this mode to store the partial sum of $V[i]^2$ (V is the input vector) 
which are computed by the multipliers and adders in EXE1 and EXE2 stages.
\subsection{Support for Distribution Representation and Comparison}
\label{sec_op_kstest}

A novel contribution of this work is to show that the statistical test for comparing empirical probability distributions can be done efficiently using the multipliers and adders already needed for the ML/DL algorithms. To the best of our knowledge, we are the first to describe the following simple and efficient hardware support for comparing PEDs and the KS test.

\begin{figure}[ht]
    \centering
    \includegraphics[width=\linewidth]{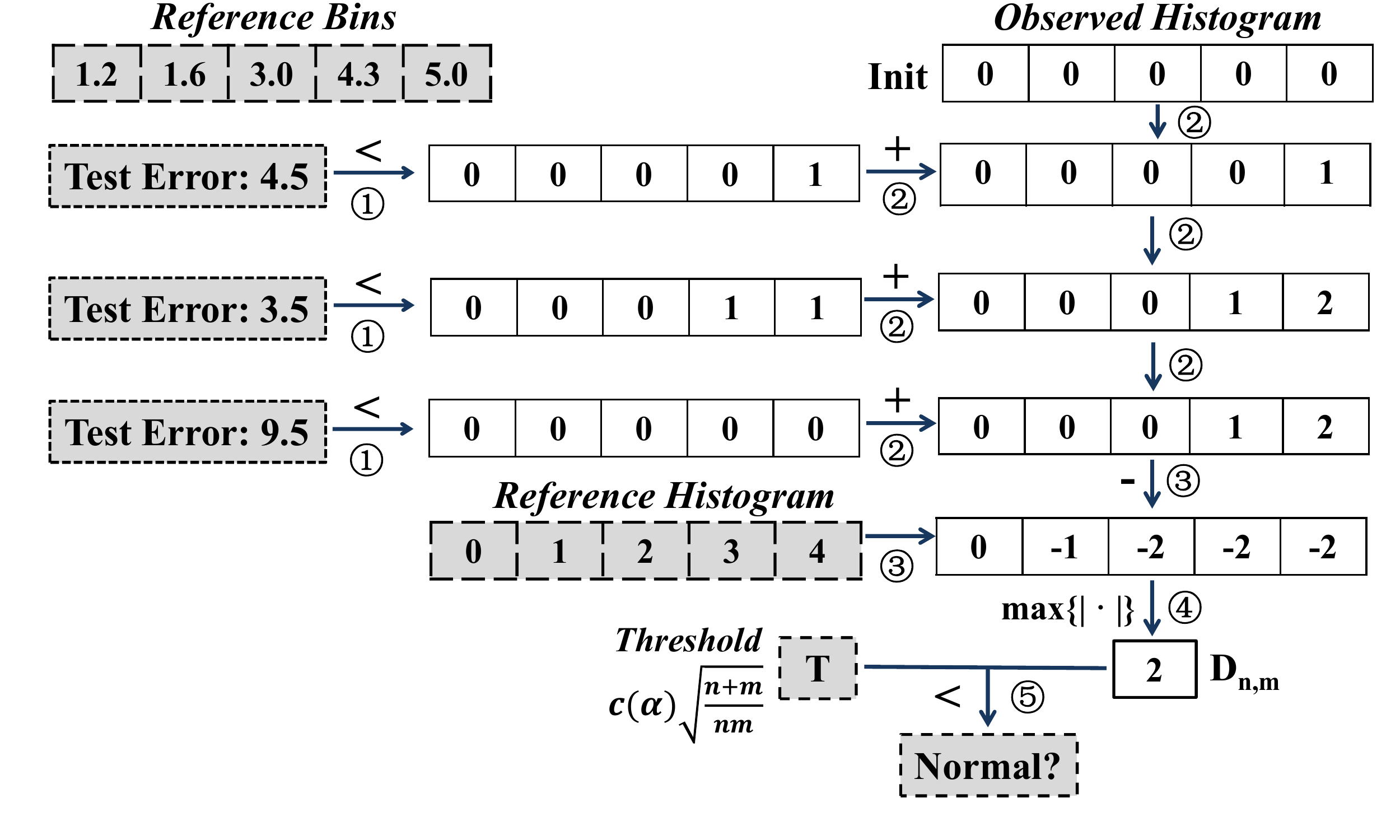}
    \caption{An example of five-step KS test. Only two new operation primitives, \textbf{VSsgt} and \textbf{Vmaxabs}, are implemented in step \ding{173} and \ding{175} to support KS test.}
    \label{fig:ks_test}
\end{figure}

We implement the KS test using a five-step workflow and show an example in \gyreffig{fig:ks_test}. The grey dotted boxes represent inputs, including the reference prediction error distribution (PED) and the observed PED and the output. The reference PED is collected in the training phase and is represented by reference bin boundaries and a cumulative histogram. 
The observed RED is collected online and represented by a series of observed errors. 

Step \ding{172}: compare an observed error with bin boundaries. The output of this step is a vector of ``0''s and ``1''s. ``1'' represents that the corresponding bin boundary is greater than the observed error, and ``0'' otherwise. This step requires a new operation for vector-scalar comparison (\textbf{VSsgt} described in~\gyrefsec{sec_parallel_operation}). Step \ding{173}: accumulate all binary vectors from \ding{172}. The accumulated vector, namely ``observed histogram'', conceptually represents the cumulative histogram of the observed errors using the reference bins. Step \ding{174} and step \ding{175}: find the largest difference in the reference and observed histograms, i.e. the KS statistic $D_{n,m}$ in \gyrefeq{equ:KSstatistic}. Step \ding{174} is a vector subtraction and step \ding{175} is a new operation, \textbf{Vmaxabs} (described in~\gyrefsec{sec_parallel_operation}), to find the maximum absolute value in a vector. Step \ding{176}: compare $D_{n,m}$ with a threshold, which is treated as a single-entry vector, 
to determine normality. 
Hence, we only need two extra operation modes, \textbf{VSsgt} and \textbf{Vmaxabs}, to support KS testing.


\section{Evaluation}
\label{sec_evaluation}


\subsection{Model Performance and Trade-offs}
\label{sec_model_perf}

\begin{figure}[ht]
    \centering
    \includegraphics[width=\linewidth]{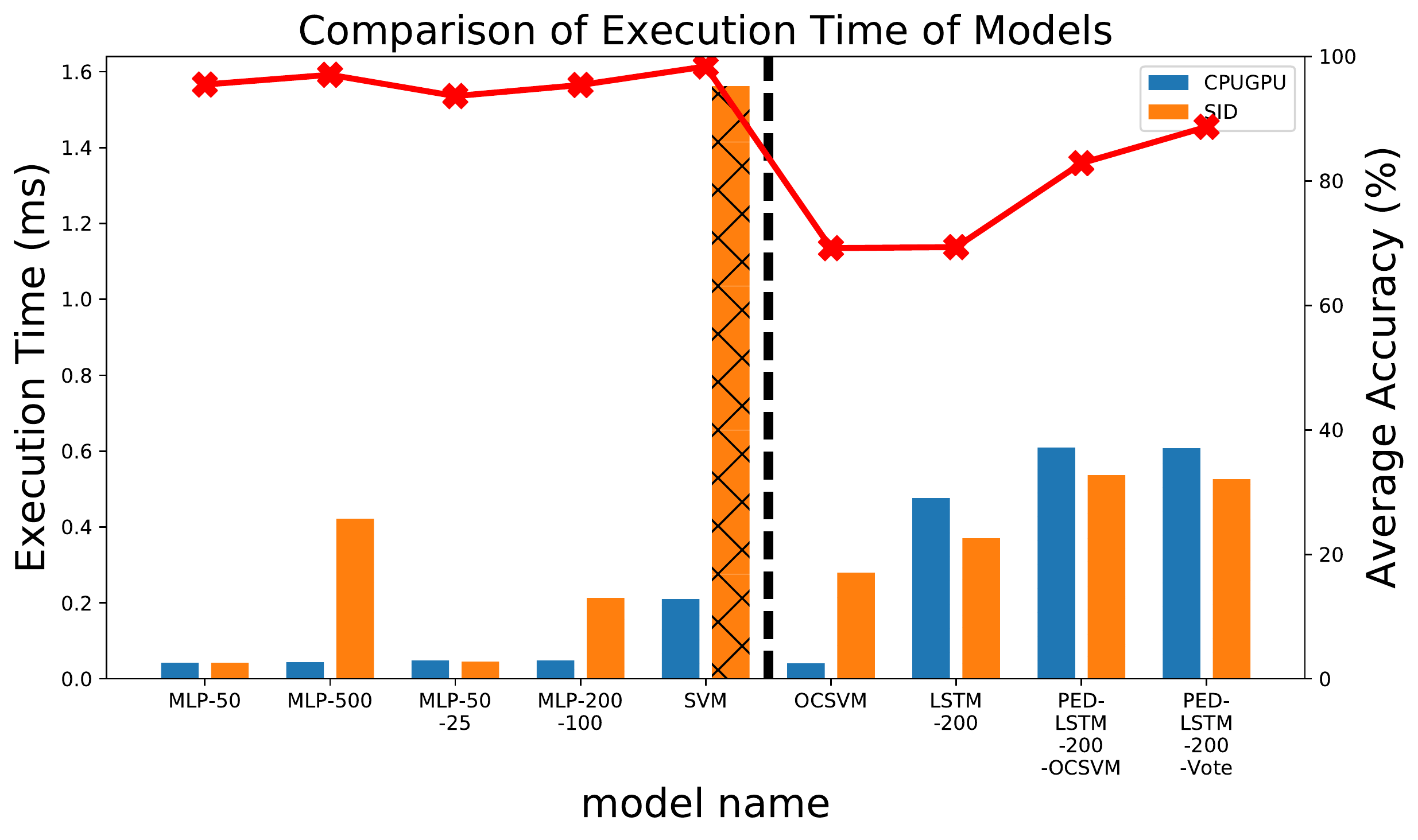}
    \caption{Execution time and accuracy of models}
    \label{img_hw_perf}
\end{figure}

\gyreffig{img_hw_perf} compares different machine learning and deep learning algorithms for their trade-offs between execution time (orange bars with crosshatch) and accuracy (red line) on the~\modulename~module and the CPU-GPU platform. The models to the left of the black dashed line are used in the IDaaS scenario when sensor data from other users are available. The models trained without the other users' data in the LAD scenario are to the right of the dashed line. 
Although the SVM algorithm achieves slightly higher accuracy than MLP-200-100 (98.4\% versus 97.1\%), it needs significantly longer execution time.

For the one-class models in the LAD scenario, we find that the KS test technique increases the detection accuracy, but also needs additional execution time. Compared to the \textbf{LSTM-200} model which involves only LSTM and error computation and has low accuracy, \textbf{PED-LSTM-200-OCSVM} increases the execution time on \modulename~by 44.9\%, to which KS test contributes 41.7\% and the following one-class SVM contributes the remaining 3.2\%. \textbf{PED-LSTM-200-Vote} increases the execution time by 41.8\%, compared to the basic LSTM model, and the execution time is mostly spent in computing the KS test. The majority vote part only contributes less than 0.1\%.

To achieve real-time detection of an impostor, the total execution time should be less than the period of sensor reading (PoS). We also tried a CPU-only platform with 32 Intel Xeon E5-2667 cores, it fails to meet this requirement. Running a single prediction on one sensor sample with the \textbf{LSTM-200} model takes more than 20 ms, i.e. the PoS of our dataset for the smartphone sensor sampling rate of 50Hz. The \modulename~module and our \textbf{CPU-GPU} platform which has a NVIDIA GTX 1080Ti GPU can meet the requirement to provide enough performance, however, the latter consumes much more energy as we will show in~\gyrefsec{sec_eval_energy}.

\subsection{Model Memory Usage}

\begin{figure}[ht]
    \centering
    \includegraphics[width=\linewidth]{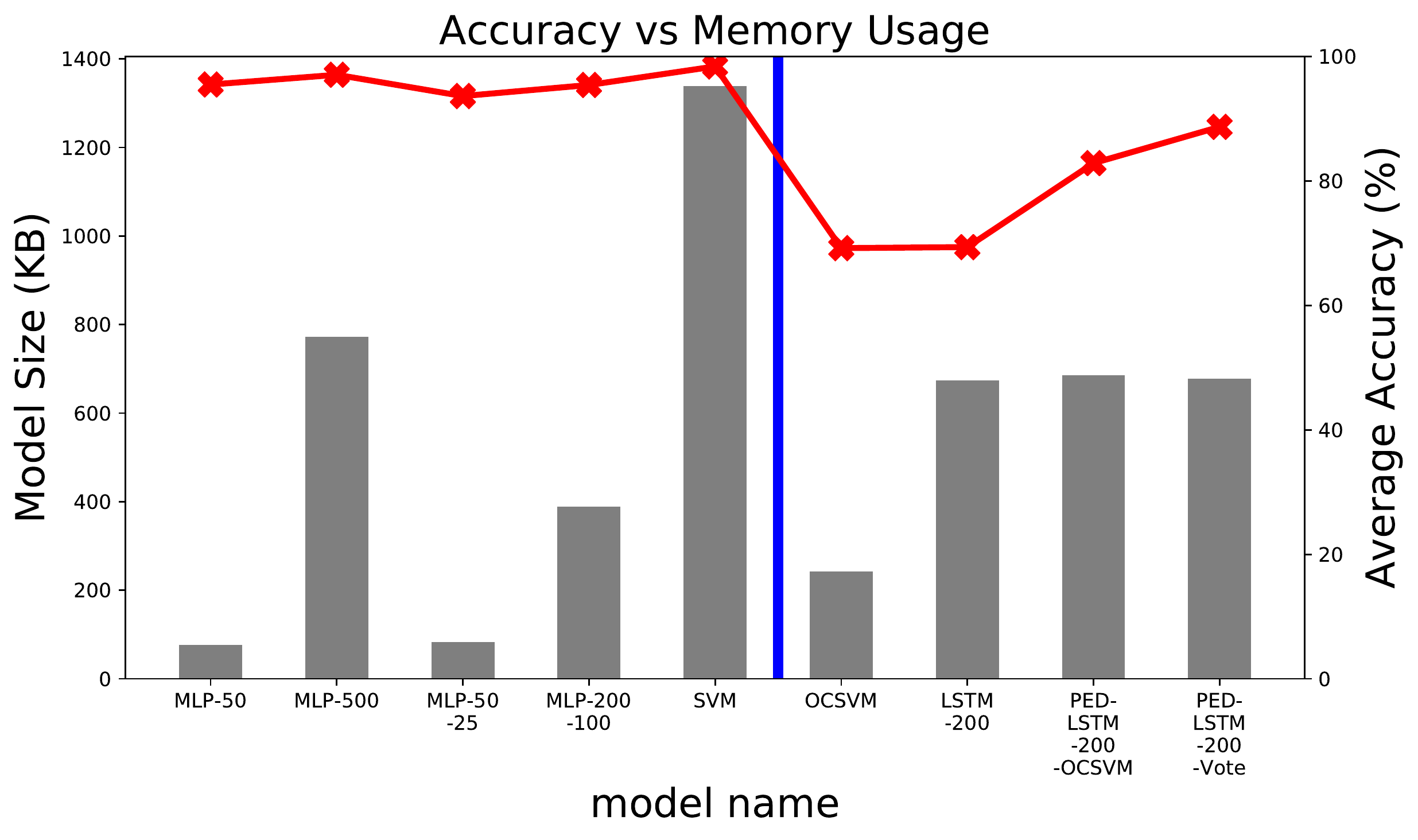}
    \caption{Number of parameters and accuracy of models}
    \label{img_hw_mem}
\end{figure}

\gyreffig{img_hw_mem}
compares models used in the IDaaS and LAD scenarios, in terms of their accuracy and the size of the model parameters. The red line stands for the average accuracy. In the IDaaS scenario, we see that a 2-layer MLP-200-100 can achieve a slightly higher detection accuracy with a smaller model size than MLP-500.
The SVM model has little improvement on the accuracy over MLP-200-100 but incurs the highest cost in terms of memory usage as it has to store many support vectors. Hence, MLP-200-100 appears to be the best for the cost (execution time + memory usage) versus accuracy trade-off.

In the LAD scenario which preserves the sensor data privacy, the standard LSTM-200 solution requires more space than the OCSVM model but gives a similar accuracy of 69\%. However, The LSTM-based models enhanced with KS-test, PED-LSTM-200-OCSVM and PED-LSTM-200-Vote, take 1.8\% and 0.6\% more space, respectively, but can significantly improve the overall accuracy. They are better choices in the cost-versus-accuracy trade-off as the improvement in accuracy is significant, and a user can choose one or the other based on his/her need for better security or usability (\gyrefsec{sec_alg_eval}).

\subsection{Hardware Design Complexity}
\label{sec_hardware_design_cost}

\begin{table}[ht]
\centering
\resizebox{0.48\textwidth}{!}{
\begin{tabular}{|l|l|l|l|}
\hline
                & \multicolumn{1}{c|}{DeltaRNN}                                                     & \multicolumn{1}{c|}{C-LSTM}                                    & \multicolumn{1}{c|}{\modulename}                                                                       \\ \hline
Functionality   & \begin{tabular}[c]{@{}l@{}}Supports Gated \\ Recurrent Unit\\ only\end{tabular}   & \begin{tabular}[c]{@{}l@{}}Supports LSTM\\ only\end{tabular}   & \begin{tabular}[c]{@{}l@{}}Supports LSTM\\ and other RNNs,\\ ML models and\\ statistical tests\end{tabular} \\ \hline
Platform        & \begin{tabular}[c]{@{}l@{}}Xilinx Zynq-7000\\ All Programmable\\ SoC\end{tabular} & \begin{tabular}[c]{@{}l@{}}Alpha Data's\\ ADM-7V3\end{tabular} & \begin{tabular}[c]{@{}l@{}}Xilinx Zynq-\\ 7000 SoC ZC706\\ Evaluation Kit\end{tabular}                      \\ \hline
FPGA            & Kintex-7 XC7Z100                                                                  & Xilinx Virtex-7                                                & \begin{tabular}[c]{@{}l@{}}XC7Z045\\ FFG900\end{tabular}                                                    \\ \hline
Design Tool     & Vivado 2017.2                                                                     & Xilinx SDx 2017.1                                              & Vivado 2016.2                                                                                               \\ \hline
Quantization    & 16-bit fixed point                                                                & 16-bit fixed point                                             & 32-bit fixed point                                                                                          \\ \hline
Slice LUT       & 261,357 (31.52X)                                                                  & 406,861 (49.07X)                                               & 8,292 (1X)                                                                                                  \\ \hline
Slice Flip-flop & 119,260 (31.40X)                                                                  & 402,876 (106.07X)                                              & 3,798 (1X)                                                                                                  \\ \hline
DSP             & 768 (48X)                                                                         & 2675 (167.19X)                                                 & 16 (1X)                                                                                                     \\ \hline
BRAM            & 457.5 (0.94X)                                                                     & 966 (1.98X)                                                    & 489 (1X)                                                                                                    \\ \hline
Clock Freq      & 125MHz                                                                            & 200MHz                                                         & 115MHz                                                                                                      \\ \hline
Power (W)       & \begin{tabular}[c]{@{}l@{}}Static: 7.9\\ Running: 15.2\end{tabular}               & Running: 22                                                    & \begin{tabular}[c]{@{}l@{}}Static: 0.12\\ Running: 0.62\end{tabular}                                        \\ \hline
\end{tabular}
}
\caption{Hardware utilization and power compared to performance-oriented RNN accelerators}
\label{tbl_hw_util}
\end{table}

We implement an FPGA prototype of the~\modulename~module. The implementation has four parallel tracks and a 256-byte scratchpad. The size of the data RAM is 1.75MB ($448 \times 4kB$ BRAM blocks) and the size of the instruction RAM is 128KB ($32 \times 4kB$ BRAM blocks). We use 32-bit fixed-point numbers, since prior work~\cite{pudiannao:Liu:2015:PPM:2694344.2694358} has shown significant accuracy degradation with 16-bit numbers. The platform board is Xilinx Zynq-7000 SoC ZC706 evaluation kit. The hardware implementation is generated with Vivado 2016.2. 

In~\gyreftbl{tbl_hw_util}, we compare \modulename~to two FPGA implementations of RNN accelerators, DeltaRNN~\cite{gao2018deltarnn} and C-LSTM~\cite{wang2018clstm}, as the LSTM or GRU RNN computation is used in the LAD scenario and accounts for most of its execution time and memory usage. DeltaRNN ignores the minor change in the input of gated recurrent unit (GRU) RNN to reuse the old computation result and thus reduces the computation workload. C-LSTM shows that, some matrices in LSTM can be represented as multiple block-circulant matrices and reduces the storage and computation complexity. These two accelerators are capable of inferring the RNN but lack the support for generating and comparing empirical PED's, which we have shown is indispensable to achieve acceptable accuracy.

The FPGA resource usage of Slice LUTs, Slice Flip-flops and DSPs of \modulename~are one or two orders of magnitude less than the other two, which shows a major difference between the minimalist \modulename~module and the performance-oriented accelerators. We measure the FPGA power consumption using TI Fusion Digital Power Designer tool. The power consumption is an order of magnitude less, making it more suitable for a smartphone. 
We use an FPGA implementation as a prototype of \modulename~to compare with existing FPGA accelerators. Further power reduction is achievable using an ASIC implementation in real smartphone products.

\subsection{Energy Consumption versus CPUGPU}
\label{sec_eval_energy}

We evaluate the energy consumption of platforms supporting the impostor detection algorithms within one period of sensor reading (PoS), which is also the period of real-time detection. The energy consumption includes that consumed during execution when the devices are actively running and also the energy when they finish execution and become idle. Specifically, the equations to model the energy consumption of the~\textbf{CPUGPU} platform and the \modulename~platform are given in~\gyrefeq{eq_detection_energy_platforms}. $P_{run}$ and $P_{idle}$ are the power consumption when a device is running the detection algorithms or in the idle state. $t_*$ is the time when the corresponding device is executing the algorithm. $T$ is the PoS, so $T-t_*$ is the time within a PoS when the corresponding device is not executing anything.
\begin{equation}
\resizebox{0.9\columnwidth}{!}{
$
\begin{split}
    E(CPUGPU)&=P_{run}(CPU) \times t_{CPU} + P_{idle}(CPU) \times (T-t_{CPU}) +\\
             &~P_{run}(GPU) \times t_{GPU} + P_{idle}(GPU) \times (T-t_{GPU}) \\
    E(\modulename) = &P_{run}(\modulename) \times t_{\modulename} + P_{idle}(\modulename) \times (T-t_{\modulename})
\end{split}
$
}
\label{eq_detection_energy_platforms}
\end{equation}
We then apply our model to compare the energy consumption used by the two platforms within a single period of sensor reading. We present a conservative comparison by not including the CPU energy in the equation of~\textbf{E(CPUGPU)} as the CPU typically runs many processes and the statistical computations it does for impostor detection may not add much to the total energy consumption.

\begin{table}[h]
\centering
\begin{tabular}{|l|c|c|}
\hline
     & Idle Power (W)
     & Running Power (W)
     \\ \hline
GPU  & 8                                                        & 56                                                          \\ \hline
\modulename & 0.12                                                     & 0.62                                                        \\ \hline
\end{tabular}
\caption{Idle and running power of different devices}
\label{tbl_power_platforms}
\end{table}

\begin{figure}[ht]
    \centering
    \includegraphics[width=0.95\linewidth]{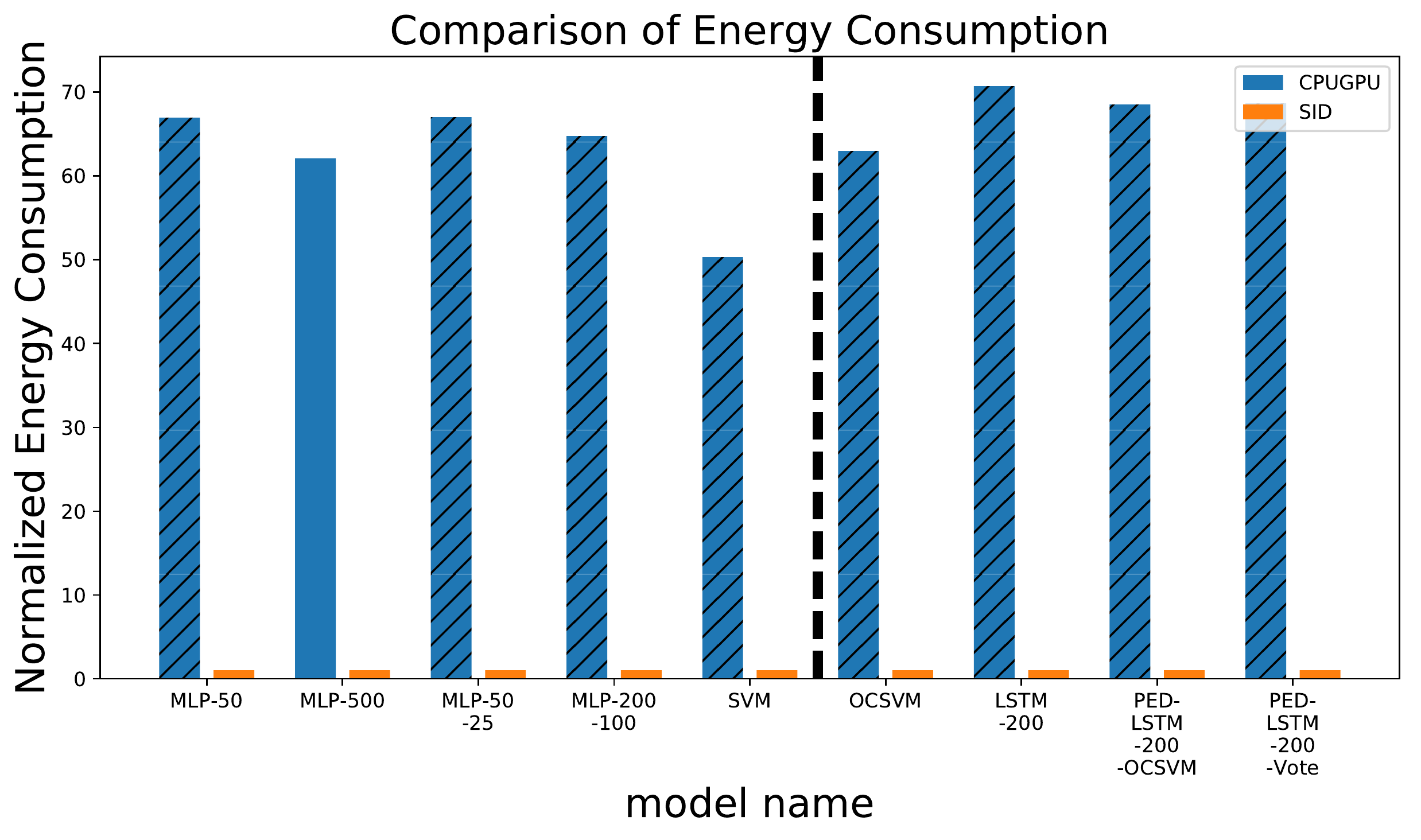}
    \caption{Normalized energy consumed by the impostor detection algorithms within one sensor reading period}
    \label{img_0605_energy}
\end{figure}

The idle and running power of GPU and~\modulename~are given in~\gyreftbl{tbl_power_platforms}. 
The GPU power is measured using the nvidia-smi tool when the GPU is idle or is running the detection algorithm. The power consumption of our \modulename~module is as given in~\gyreftbl{tbl_hw_util}.
\gyreffig{img_0605_energy} shows the energy consumption normalized to the \modulename~platform when running the models on the two platforms. 
The average consumption for all models on the CPU-GPU platform is 64.41X 
higher than that of the \modulename~platform. Interestingly, the ratio of energy consumption is close to the ratio of idle powers of the two platforms ($\frac{8}{0.12}=66.66$). The takeaway is that when the execution time is short compared to the PoS (\gyrefsec{sec_model_perf}), the energy consumption is determined by that consumed in the idle state, which complies with our effort to reduce hardware cost and static power. \modulename~also frees up the GPU for its original graphics purposes, rather than tying it up for impostor detection.

\section{Related Work}
\label{sec_related_work}



Dedicated accelerators for a single machine learning (ML) algorithm have been built by researchers, 
e.g. for SVM \cite{fpga:svm:cascade}\cite{svm:training:fpga:papadonikolakis2010heterogeneous}, for k-th neatest neighbors~\cite{knn:acceleration:schumacher2008hardware} and for k-means\cite{accelerator:k-means}.

Accelerators are also built for deep learning (DL) algorithms, e.g. for deep neural networks (DNNs). 
Many of these accelerators are designed with insights identified in software execution. EIE~\cite{han2016eie} and Minerva~\cite{reagen2016minerva} exploit data sparsity of weights and activations during inference to improve performance and energy efficiency. The sparsity in training is exploited in~\cite{rhu2018compressingdma} to improve performance. Minerva~\cite{reagen2016minerva} presents a framework to reduce power consumption by finding the optimal data quantization in the accelerator with software exploration. 
Weight sharing in CNN is identified by~\cite{song2018insituai} in early software exploration before designing a dedicated accelerator. These accelerators benefit from different properties of DNN models to improve performance or energy efficiency, but the versatility for implementing other ML/DL algorithms is not considered, as we do for \modulename.

Some hardware accelerators also target supporting multiple machine learning models.~\cite{multi:cnn:kmeans:accleration:majumdar2011energy} evaluates the acceleration of four models in an embedded CPU-GPU-Acce\-lerator system. MAPLE~\cite{multi:maple:majumdar2012massively} accelerates the vector and matrix operations found in five classification workloads. PuDianNao~\cite{pudiannao:Liu:2015:PPM:2694344.2694358} highlights the non-vector operations and data locality in seven ML techniques. While these work support their chosen ML/DL algorithms, they do not support other needed processing such as statistical KS tests as we do. 

For minimalist hardware design, we incorporate conventional energy-saving techniques, e.g. the tiling method~\cite{diannao:Chen:2014:DSH:2541940.2541967}, that can benefit multiple ML/DL algorithms. We do not implement hardware modules for a specific ML/DL algorithm. To the best of our knowledge, we are the first to explore the direction of delivering versatility and sufficient hardware performance to solve a critical security problem, like smartphone impostor detection, with reduced energy consumption, rather than shooting for maximum performance or minimum power efficiency. 

\section{Conclusions}
\label{sec_conclusion}

We show how sensors in a smartphone can be used to detect smartphone impostors and theft by identifying the current user using Deep Learning models like MLP and LSTM or GRU and the empirical PED's. We explore the algorithms that are needed in two scenarios, IDaaS and LAD, when users can or cannot use other people's data for training.

We design a hardware module, \modulename~, to support the best impostor detection algorithms we found in both scenarios. It is also versatile enough to support other ML/DL algorithms as well as collecting and comparing empirical probability distributions that we use to represent user behavior. This enables users of \modulename~to tradeoff security with data privacy in choosing one of the two scenarios, as well as choosing trade-offs in security with usability and accuracy with execution time and memory storage. Our \modulename~macro instructions can also map the same code to \modulename~hardware substrates with different amounts of parallel computation tracks. For efficiency, \modulename~reduces hardware resource usage by maximally reusing functional units. 
Our evaluation shows that our FPGA implementation of \modulename~has a major difference with accelerators targeting similar ML/DL models: \modulename~provides sufficient performance with minimal hardware and energy costs, 
which are one to two orders of magnitude less than performance-oriented FPGA accelerators. Using a general energy consumption model, we also show that the consumption of the FPGA implementation of a~\modulename~module is 64.41X smaller than the CPU-GPU platform.

We hope to have shown a new direction for computer architecture in the ML/DL domain: consider broader goals and trade-offs for real security (or other domain) needs rather than focus on optimizing just performance or power. Also, we hope to have shown the importance of a design methodology that considers both algorithm and architecture optimizations in solving critical problems like impostor detection, preventing subsequent confidentiality and integrity breaches.

\begin{appendices}

\section{Classification Algorithms}
\label{apd_class_alg}

\bheading{Logistic Regression} uses a logistic function to model a binary dependent variable (impostor or not), i.e.:
\begin{equation}
\label{eq_linearreg}
  \hat y = f(w^T x + b)
\end{equation}
where $x$ denotes the input, $w^T$ represents the transpose of $w$, and $f(z) = \frac{1}{1+e^{-z}}$ is the logistic function. The model prediction $\hat y \in (0,1)$ can be interpreted as the probability of $x$ being an impostor. The model parameters $w$ and $b$ are trained to fit the training data by minimizing the cross-entropy loss.



\bheading{Support Vector Machine} is a popular linear model in machine learning. To perform a binary classification, it identifies a hyperplane ($w^T x + b$ in \gyrefeq{eq:SVM:inner}) as the decision boundary 
where $w$ is the normal vector of the hyperplane. $Sign(\cdot)$ is the sign function. Unlike the logistic regression, $w$ and $b$ are trained by maximizing the hinge loss. 
A property of the SVM is $w$ can be represented by a weighted average of support vectors (a subset of data points in the training set).
\begin{align}
  \hat y = sign(w^T x + b) 
   = sign(\sum^{N}_{i=1} a_i \langle v_i, x \rangle + b)
\label{eq:SVM:inner}
\end{align}
where $N$ is the number of support vectors, $v_i$'s are the support vectors, $a_i$'s are the weights in representing $w$ of support vectors and $< \cdot >$ is the inner product of vectors.


If the data points are not linearly separable in the original space, a typical way is to map the data into a high-dimensional space where they can be linearly separated. 
A kernel function can be leveraged 
to obtain an equivalent classifier in the high-dimensional space, by replacing the inner product in \gyrefeq{eq:SVM:inner}. 
The kernel function used in this paper is the Gaussian Kernel, i.e. $ \kappa(u,v)=e^{-\gamma \norm{u-v}^2}$.

\bheading{Kernel Ridge Regression.} Similar to SVM, ridge regression can be used as a linear classifier for binary classification:
\begin{align}
    \hat y = sign(w^Tx + b)
\end{align}

However, different from SVM and logistic regression, ridge regression is a technique specifically designed to deal with ill-posed problems with very limited data, by introducing an extra $l_2$ norm of $w$ in the loss function during training. Kernel ridge regression (KRR) is when the same kernel trick as for SVM can be applied to ridge regression.

\bheading{Multi-layer Perceptron (MLP)} is a family of feed-forward neural network models, consisting of an input layer, an output layer and one or more hidden layers in between. Each layer 
of neurons linearly transforms $all$ signals from the previous layer, applies a non-linear activation function and outputs to the next layer. Finally, a softmax function \footnote{Softmax($z$) = $ \frac{e^{z_i}}{\sum_i e^{z_i}}$} is applied to output the probabilities of classes. Formally,
\begin{align*}
    h_{1} &= f(W_1^T x+b_1) \\
    &... \\
    h_{n} &= f(W_{n}^T h_{n-1}+b_{n}) \\
    \hat y &= softmax(h_n)
\end{align*}

where $h_i$ denotes the output of layer $i$, $f$ denotes a non-linear activation function, e.g. sigmoid or ReLU \footnote{ReLU($z$) = $z$ if $z\ge 0$, else 0}. 
Similar to logistic regression, the weight-matrix $W$'s and the bias $b$'s are trained by minimizing the cross-entropy loss.




\end{appendices}

\newpage


\bibliographystyle{IEEEtranS}
\bibliography{refs}

\end{document}